\documentclass[10pt,preprint]{aastex}

\begin{document}

\title{A Cluster of High Redshift Quasars with Apparent Diameter 2.3 Degrees}

\author{H. Arp}

%\offprints {H. Arp}

\affil{Max-Planck-Institut f\"ur Astrophysik, Karl Schwarzschild-Str.1,
  Postfach 1317, D-85741 Garching, Germany}
 \email{arp@mpa-garching.mpg.de}

\author{C. Fulton}
\affil{Centre for Astronomy, James Cook University, Townsville
Queensland, 4811, Australia}
\email{mainseq@thevine.net}

%\date{Received}

\begin{abstract}

During analysis of the relation of quasars to galaxies in the 2dF deep field a concentration of quasars was noted. Most striking was the closeness in redshift of 14 quasars about the mean redshift $z = 2.149$ with a range of $\pm .018$. The cluster in spite of its high redshift subtends an area of diameter more than $2.3$ degrees on the sky. At conventional redshift distance its diameter would be $181$ mega parsecs and the back should be receding with about $10,000\ km/s$ with respect to the front.

\end{abstract}


\section{Introduction}

In the two 2dF surveys, the 2dF Galaxy Redshift Survey (2dFGRS)
examines $\sim250,000$ galaxies \citep{sadler2002} and the 2dF Quasar
Redshift Survey (2QZ) examines $\sim25,000$ quasars.
\citep{croom2001}. Together they furnish a vast wealth of redshift data. The 2dF deep field is particularly rich because it is the densest quasar data set to date and because the 2dF objects in the field can be supplemented with extragalactic objects from other surveys. Detailed study of this deep field data set has revealed striking attributes of a number of galaxy systems including a very prominent one reported in this paper.

\subsection{Excess Object Densities in the 2dF Deep Field}

\label{S-ExcessObjectDensitiesInThe2dFDeepField}

As originally envisioned the 2QZ was to cover two strips on the sky each of $\sim5$ hours in right ascension by $5^{\circ}$ in declination, for a total area of $750^{{\circ}^2}$ \citep{croom2001}, but extensions and contractions were made at the strip ends during the observation program, yielding a final 2QZ catalogue covering $740^{{\circ}^2}$ \citep{outram2003}. When all redshift-available objects are downloaded from NED in the entire 2QZ area, duplicate objects removed, and the respective 2dF survey magnitude limits applied to the galaxies and quasars, the resulting collection of objects can be analyzed to obtain the average numbers of galaxies and quasars per square degree as shown in Table \ref{T-AverageNumbersOfObjectsInThe2dFDeepField}. The \textit{subject} count records the occurrence of galaxies and quasars inside a circle of radius $30'$ around each galaxy and the \textit{background} count records the occurrence of galaxies and quasars in a concentric annulus of equal area enclosing the subject circle. The \textit{excess} count is obtained by subtracting the background count from the subject count. On average we can expect $33.1$ quasars per square degree within a $30'$ radius of each galaxy, or an average of

\begin{displaymath}
33.1 \times \pi \times ({.5}^{\circ})^2 = 26
\end{displaymath}
quasars around each putative parent galaxy.

\begin{table}[ht]
\begin{center}

\caption{Average Numbers of Objects in the 2dF Deep Field}

\label{T-AverageNumbersOfObjectsInThe2dFDeepField}

\begin{tabular}{rrrrrr} \\
{\bfseries Object\/}& {\bfseries   Subject Count\/} & {\bfseries Background Count\/} & {\bfseries    Excess Count\/} \\
{\bfseries   Type\/}& {\bfseries $/^{{\circ}^2}$\/} & {\bfseries  $/^{{\circ}^2}$\/} & {\bfseries $/^{{\circ}^2}$\/} \\
\hline

Galaxies            &                       $164.3$ &                        $145.1$ &                       $ 19.3$ \\
Quasars             &                       $ 33.1$ &                        $ 30.5$ &                       $  2.7$ \\

\end{tabular}

\end{center}
\end{table}

\subsection{Criteria of Physical Association}

\label{S-CriteriaOfPhysicalAssociation}

\citet{fulton2007}(paper I) have analyzed the positions, redshifts, and magnitudes of $\sim118,000$ galaxies and $\sim25,000$ quasars in the 2dF deep field. The examination of individual samples revealed concentrations of high $z$ galaxies and quasars near galaxies. A natural extension of the analysis was to determine the average densities of objects over the survey area as a whole. One method is to derive densities around galaxies and let their approach to constancy with increasing radius define the average background. An alternative method, outlined in Section \ref{S-ExcessObjectDensitiesInThe2dFDeepField}, yields a \textit{minimum} excess. When applied to the 2dF deep field this method demonstrates the excess unequivocally, strongly indicating physical association of high redshift ($z > .5$) objects with low redshift ($z < .3$) galaxies.

The main analysis of paper I used a custom algorithm to detect families of quasars associated with parent galaxies. Knowledge of the galaxy and quasar excess was not used directly by the algorithm. Rather, the existence of the excess served as one of the empirical clues that drove the formulation of the algorithm. The algorithm confirmed, as claimed by \citet{arp1998} (p.243, 285), that the redshifts of the quasars associated with galaxies tend to decrease as their separation from their parent galaxy increases. The algorithm also confirmed a subsequent assertion by \citet{arp2005} that the quasar redshifts preferentially fall near the periodic Karlsson peaks when viewed from the rest frame of the galaxy.

A key feature of the paper I detection algorithm is a series of statistical calculations that measure the significance of each test result. One of these is a Monte Carlo control trial edifice that employs the same logic and data inputs as the test under statistical scrutiny. This control differs only in that it picks redshifts randomly from the entire population of redshifts in the 2dF deep field and, prior to performing a control trial, substitutes these random redshifts for the actual redshifts of candidate companions in the $30'$ circle encompassing the putative parent galaxy. This test and control trial ritual was performed with various derived constraints. The applied constraints first ensured that the quasars fell significantly close to the parent galaxy, then that they declined in redshift with increasing separation from the parent, and finally that they fell close to a Karlsson redshift peak (\citet{karlsson1971}; \citet{karlsson1973}; \citet{karlsson1977}; \citet{karlsson1990}.) The significance of the physical associations increased with each succeeding constraint. The significance of any given test rose as the number of quasars required for detection of a family was increased. When a minimum of four high redshift companions was required for family detection, the number of control trial detections dropped dramatically. For the most constrained test case the control became negligible. It is the resulting catalogue of quasar families in which the quasar cluster reported herein was encountered.

\subsection{Among the 44 Most Probable Associations - AM2230-248}

\label{S-AmongThe44MostProbableAssociations-AM2230-248}

In order to work with a manageable number of cases that would allow individual inspection the program of paper I was asked to excerpt from the most constrained test a list of the families with the largest number of detected companions. The list supplied $44$ galaxies with  $7 - 9$ such companions. Glancing through these associations revealed the surprising appearance of families in which many of the quasar companions were strikingly similar in redshift. In one case the redshifts of all $7$ quasars within a radius of $d = 30'$ were closely the same.

The central galaxy was AM 2230-284 \citep{arp1987b} and offered the chance to see if the Karlsson peaks could be picked out of the raw redshifts in the field (uncorrected to the parent galaxy redshift.) Surprisingly there was an extremely sharp peak at $z = 2.149$.

\begin{figure}
\figurenum{1} \plotone{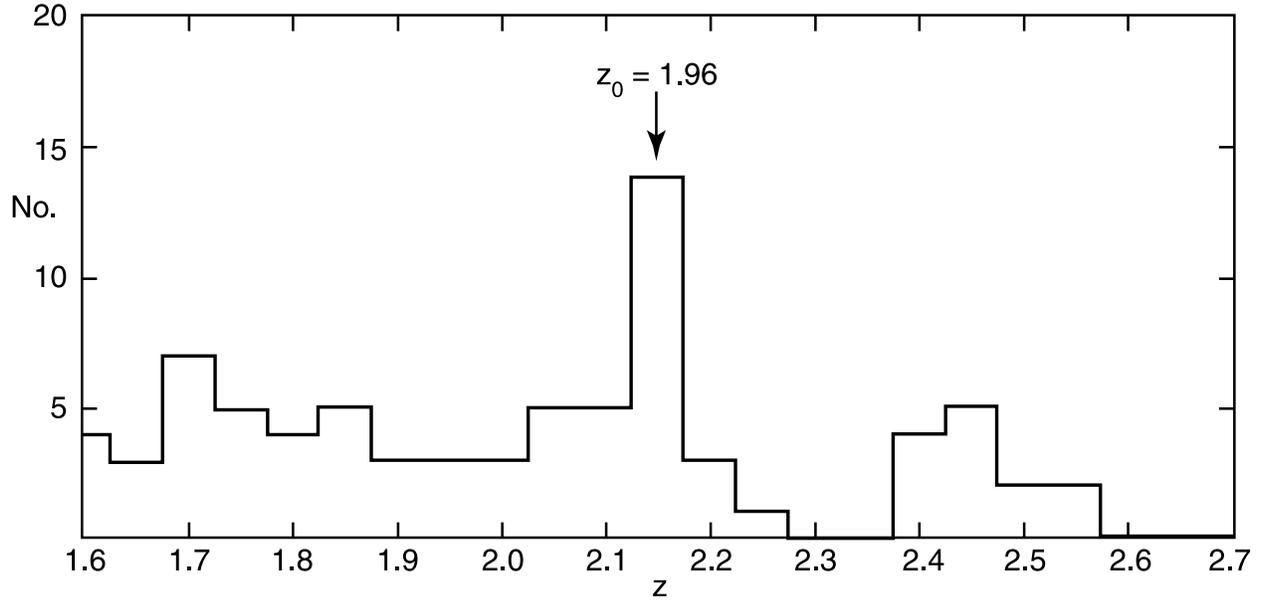}

\caption{All NED Listed Quasars within $70'$ of AM 2230-284 that have redshifts between $z = 1.6$ and $2.7$. When referenced to the parent, the
Karlsson peak of $z_0 = 1.96$ should be observed at $2.149$. The arrow points to $z = 2.150$.}

\label{F-AllNEDListedQuasarsWithin70ArcminOfAM2230-284}

\end{figure}

\begin{figure}
\figurenum{2} \plotone{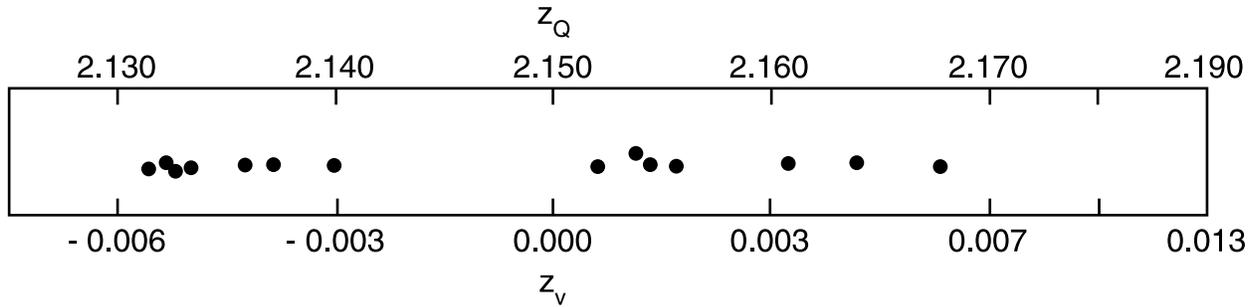}

\caption{The Redshifts ($z_Q$) of the $14$ QSOs within $70'$ of AM 2230-284. Their mean is $z = 2.148$.}

\label{F-TheRedshiftsOfThe14QSOsWithin70ArcminOfAM2230-284}

\end{figure}

After several trials it was ascertained that the densest area of quasars of this redshift was within a radius of $d = 70'$. In Figure \ref{F-AllNEDListedQuasarsWithin70ArcminOfAM2230-284} are plotted all QSOs with redshifts from $z = 1.6$ to $2.7$. Looking at the peak in the highest resolution in Figure \ref{F-TheRedshiftsOfThe14QSOsWithin70ArcminOfAM2230-284} shows it to be double with the mean value of the 14 QSOs at $z = 2.148$. The extraordinary precision of this result is what is so surprising. The periodicity model required a peak at $z = 2.149$ for the observed redshift of the parent $z_P = .064$. In fact there is a $5$ sigma signal at the predicted wavelength, i.e. the wavelength of the major Karlsson peak, $1.96$, shifted to within $.001/2.149 = .0005$ of where it should be.

\begin{figure}
\figurenum{3} \plotone{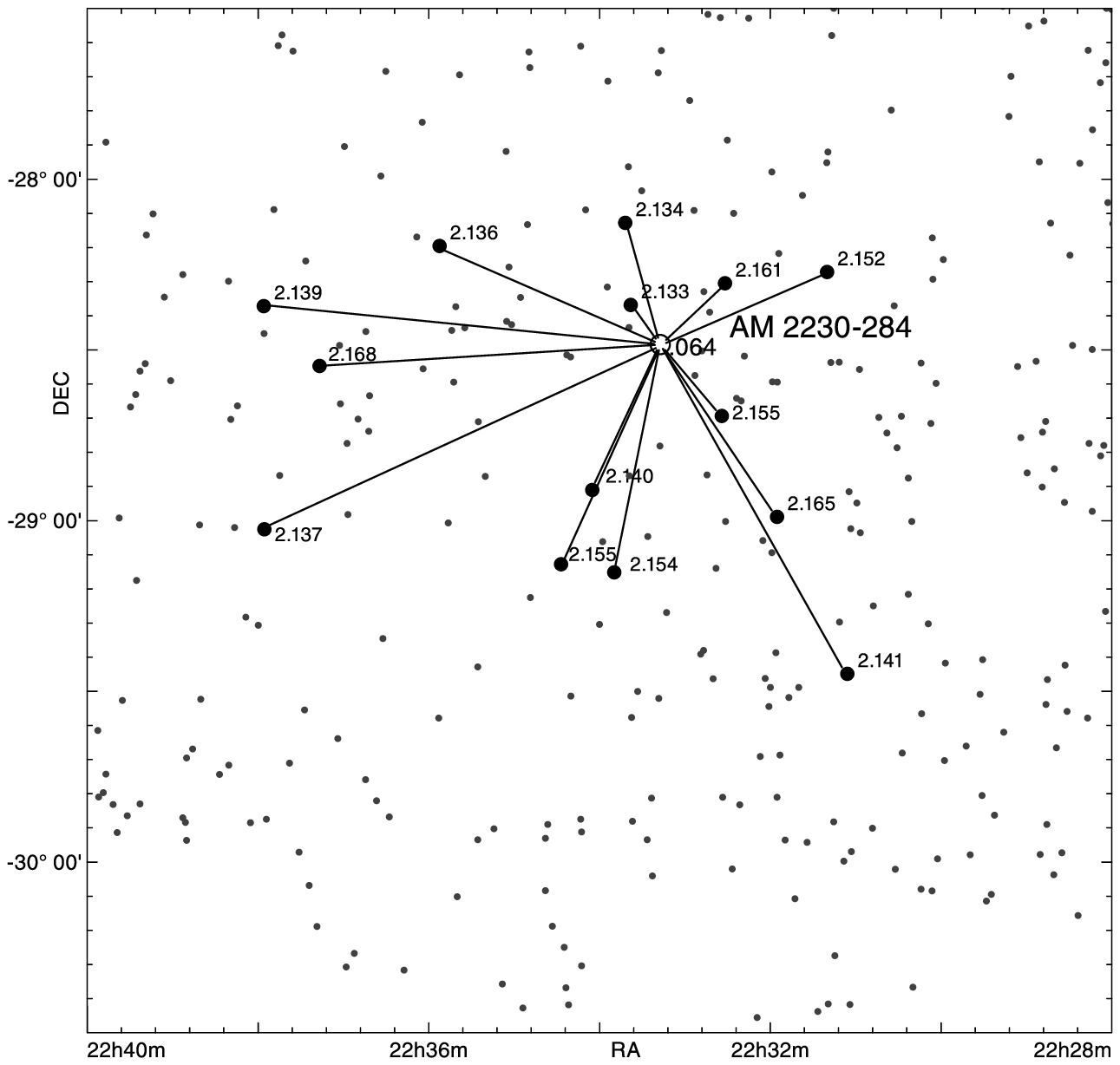}

\caption{An Unprecedentedly Large Quasar Family. All quasars in a $3 \times 3$ degree 2dF field are plotted as filled circles. Inside $70'$ and for $z$ between $2.125$ and $2.175$, the filled circles are larger and labeled with redshifts. Diagram courtesy D. Carosati.}

\label{F-AnUnprecedentedlyLargeQuasarFamily}

\end{figure}

There are specific properties of this association that are predicted from the ejection model for quasars by \citet{narlikar1993}. Briefly summarized they are:

\begin{itemize}

\item{QSOs are ejected in opposite directions conserving linear momentum. Figure \ref{F-TheRedshiftsOfThe14QSOsWithin70ArcminOfAM2230-284} shows $7$ QSOs with positive (presumably Doppler) velocity shifts and $7$ with negative shifts.}

\item{The mean approaching and receding ejection velocities are very much the same.  Extension along the lines of ejection can be slowed or deviated by moving individual QSOs around but the average usually stays closely balanced.}

\item{The parent galaxy is an Arp/Madore peculiar galaxy. It is moderately bright at $B = 17.33$ mag. Its peculiarity is its compactness (high surface brightness) usually an indicator of active physical processes.}

\item{The redshift of the parent galaxy is $z = .064$. This is very interesting because $z = .060$ is the first redshift in the Karlsson periodicity and is observed at $z = .061$ \citep{burbidge1968} and at $z = .062$ \citep{arp1990a} and in some cases up to $.065$.}

\end{itemize}

Interestingly, associated QSOs can cluster at different Karlsson peaks other than 1.96. For example the three quasar families discussed in Section 4.9 of paper I has one family distributed 5 at $z_K$ = .96 and 4 at $z_K$ = 1.41. Another family is concentrated at $z_K$ = 0.60 and 1.96. This may reflect the absolute magnitude differences in the quasars at the Karlsson peaks (see the roof relation" in \citet{arp1987a}, p.68.)

In the catalogue excerpt of 44 parents with from 7 to 9 associated quasars there are numerous examples of families with several quasars of closely similar redshifts. These need to be studied with images of the group, spectral and magnitude parameters connected to the age of the association, and the kind of parent galaxies.

The association of quasars with AM 2230-284, however, is the most striking case among the many reported by the algorithm developed in the paper I analysis. Overall it represents an active galaxy which is itself at the lowest Karlsson peak and, judging by the apparent magnitude of its associated quasars, probably much closer than its redshift distance. The fact that there are so many quasars all of nearly the same redshift around this galaxy marks them as being associated with a high degree of probability. Then the astonishingly exact numerical agreement with the strongest Karlsson peak renders the association even more significant. Further evidence is even available from the individual redshifts of the quasars. Deviations from the narrow Karlsson peak are very small but almost perfectly balanced plus and minus across the galaxy, confirming the empirical model of nearby galaxies ejecting young, compact objects that lose their intrinsic redshifts as they evolve into normal galaxies.

Of course when families contain QSOs of very similar redshifts it must
be considered whether they could belong to a distant cluster at the
conventional redshift distance in an expanding universe. Figure
\ref{F-AllNEDListedQuasarsWithin70ArcminOfAM2230-284} shows 14 QSOs
around the parent AM 2230-284. Can it be argued that this is a
background cluster of QSOs upon which accidently falls a foreground
galaxy? Aside from the evidence, statistical and interaction, obtained
in the paper I analysis there is evidence from Figure
\ref{F-TheRedshiftsOfThe14QSOsWithin70ArcminOfAM2230-284} that the
range of QSO velocities with respect to the mean is well measurable
but still not enough for expansion of such a large volume of space.

Figure \ref{F-TheRedshiftsOfThe14QSOsWithin70ArcminOfAM2230-284} shows that the seven plus and seven minus velocities yield a range in peculiar velocity, $z_v$, of

\begin{displaymath}
|-.006| + |.006| = .012
\end{displaymath}
\begin{displaymath}
.012 \times 300,000\ km/s = 3,600\ km/s
\end{displaymath}
It is interesting to calculate what the rate of expansion would be if the cluster were at its conventional redshift distance. First of all, how far away would it be? If the velocity of light is taken to be $300,000\ km/s$, then the redshift $z = 2.149$ is \citep{borner1988}

\begin{displaymath}
v / c =.817
\end{displaymath}
\begin{displaymath}
v = 245,100\ km/s
\end{displaymath}
Using the Hubble constant $H_0 = 55\ km/s/Mpc$ \citep{arp2002}

\begin{center}
$r = 4,456\ Mpc =$ the distance to the cluster

$D = 181\ Mpc =$ the diameter of the cluster
\end{center}

Hence the cluster should be expanding with $9,955\ km/s$. But only $3,600\ km/s$ is measured and most, if not all, of that is deemed ejection velocity. At the conventional redshift distance, however, just the expansion of space should imprint nearly $3$ times as much front to back expansion velocity than actually measured for this quasar cluster.

The central galaxy and its quasar family are diagramed in Figure \ref{F-AnUnprecedentedlyLargeQuasarFamily}. As unprecedentedly large as this cluster of quasars is, with the usual expanding universe assumptions it would appear to be a high priority target for photometry and spectra at fainter levels. It would seem that galaxies should be present in abundance some magnitudes fainter at closely the same redshift if redshifts are an accurate measure of distance.

\section{Acknowledgements}

This research has made use of NED which is operated by the Jet Propulsion Laboratory, California Institute of Technology, under contract with the National Aeronautics and Space Administration.

This research has made use of the Simbad data base which is operated at CDS Strasbourg, France

\end{document}